%% file: main.tex
\documentclass{egpubl}

\usepackage{sca2022}

\SpecialIssuePaper         

\usepackage[T1]{fontenc}
\usepackage{dfadobe}  
\usepackage{marvosym}
\biberVersion
\BibtexOrBiblatex
\usepackage[backend=biber,bibstyle=EG,citestyle=alphabetic,backref=true]{biblatex} 
\electronicVersion
\PrintedOrElectronic

\ifpdf \usepackage[pdftex]{graphicx} \pdfcompresslevel=9
\else \usepackage[dvips]{graphicx} \fi

\usepackage{float}
\usepackage{egweblnk} 
\input{content/_preamble.tex}

\begin{document}
\input{content.tex}
\end{document}

%% file: content/_preamble.tex
\usepackage{bm}

\newcommand{\lb}[1]{\textcolor{black}{{#1}}}
\newcommand{\yyj}[1]{\textcolor{black}{{#1}}}

\usepackage{pifont}

\usepackage{graphicx}
\usepackage{subcaption}
\usepackage{multirow}
\usepackage{amsmath}
\usepackage{amssymb}
\usepackage{soul}
\usepackage{mathrsfs}

\usepackage{xspace}
\newcommand{\ie}{\textsl{i.e.}~}

\newcommand{\keep}[1]{}
\newcommand{\old}[1]{}

\DeclareMathOperator*{\argmax}{arg\,max}
\DeclareMathOperator*{\mse}{mse}

\newcommand{\fig}{Figure{}~}
\newcommand{\eqn}{Equation{}~}

\newcommand{\vrc}{\ensuremath{{\bm{o}}}}
\newcommand{\heading}{\ensuremath{{\bm{h}}}}
\newcommand{\genstate}{\ensuremath{{\bm{g}}}}
\newcommand{\genctrl}{\ensuremath{{\bm{z}}}}
\newcommand{\genx}{\ensuremath{{\bm{x}}}}
\newcommand{\lob}{\ensuremath{\text{lo}}}
\newcommand{\upb}{\ensuremath{\text{up}}}
\newcommand{\hb}{\ensuremath{\text{sig}}}
\newcommand{\contact}{\ensuremath{{\bm{c}}}}
\newcommand{\rlstate}{\ensuremath{{\bm{s}}}}
\newcommand{\rlState}{\ensuremath{{S}}}
\newcommand{\rlaction}{\ensuremath{{\bm{u}}}}
\newcommand{\rlAction}{\ensuremath{{U}}}

%% file: content.tex

\title
      {Neural3Points: Learning to Generate Physically Realistic Full-body Motion for Virtual Reality Users}

\author[Ye et al.]
{\parbox{\textwidth}{\centering 
      Yongjing Ye$^{1,2}$\orcid{0000-0002-1027-3382},
      Libin Liu$^{3}$\thanks{Corresponding author:libin.liu@pku.edu.cn}\orcid{0000-0003-2280-6817}, 
      Lei Hu$^{1,2}$\orcid{0000-0001-8938-5071}, 
      and Shihong Xia$^{1,2}$\thanks{Corresponding author:xsh@ict.ac.cn}\orcid{0000-0001-8938-5071}
        }
        \\
{\parbox{\textwidth}{\centering $^1$Institute of Computing Technology, Chinese Academy of Sciences, China\\
$^2$University of Chinese Academy of Sciences, China\\
         $^3$Peking University, China
      }
}
}


%


\teaser{
 \includegraphics[width=0.92\linewidth]{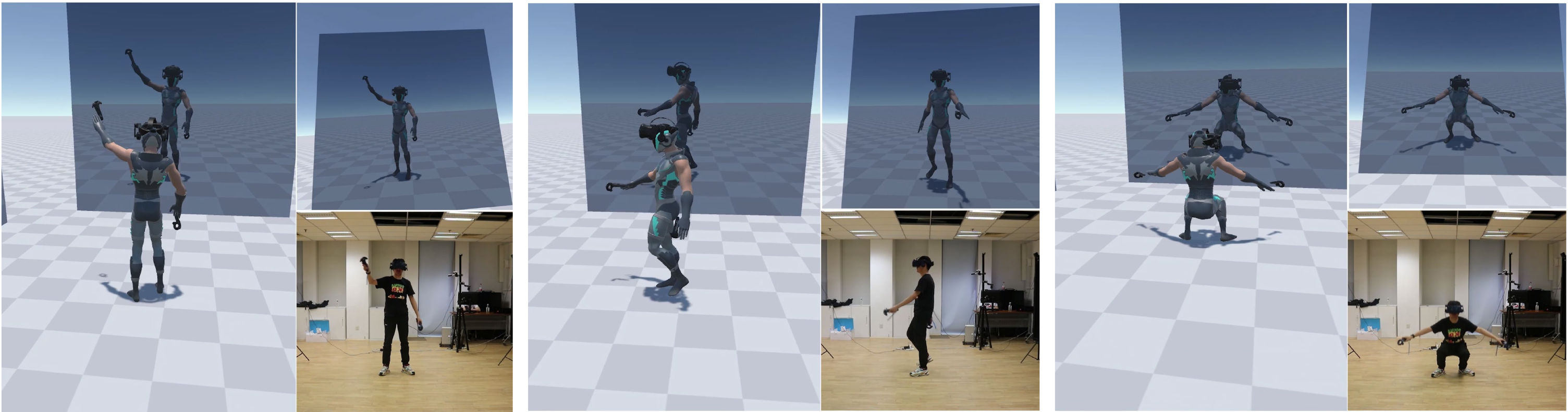}
 \centering
  \caption{We present a method for real-time full-body tracking using three VR trackers provided by a typical VR system: one HMD (head-mounted display) and two hand-held controllers. %
  The bottom-right image of each sub-figure shows a user playing with the VR system. The other images show the simulated avatar of the user in the third-person point of view (left) and the first-person point of view (top-right), respectively. Virtual mirrors are placed in front of the user, so they can see their avatar easily.}\label{fig:teaser}
}

\maketitle

\input{content/0_abstract.tex}

\input{content/1_introduction.tex}

\input{content/2_related_work.tex}

\input{content/3_system_overview.tex}

\input{content/4_control_predictor.tex}

\input{content/5_implementation.tex}

\input{content/6_results.tex}

\input{content/7_discussion.tex}

\input{content/8_ack.tex}

\printbibliography   

%% file: content/0_abstract.tex
\begin{abstract}
    Animating an avatar that reflects a user's action in the VR world enables natural interactions with the virtual environment. It has the potential to allow remote users to communicate and collaborate in a way as if they met in person.
    However, a typical VR system provides only a very sparse set of up to three positional sensors, including a head-mounted display (HMD) and optionally two hand-held controllers, making the estimation of the user's full-body movement a difficult problem. 
    In this work, we present a data-driven physics-based method for predicting the realistic full-body movement of the user according to the transformations of these VR trackers and simulating an avatar character to mimic such user actions in the virtual world in real-time.
    We train our system using reinforcement learning with carefully designed pretraining processes to ensure the success of the training and the quality of the simulation. We demonstrate the effectiveness of the method with an extensive set of examples. 

\ccsdesc[500]{Computing methodologies~Physical simulation}
\ccsdesc[500]{Computing methodologies~Virtual reality}
\ccsdesc[500]{Theory of computation~Reinforcement Learning}
\ccsdesc[300]{Computing methodologies~Motion capture}

\printccsdesc   
\end{abstract}  

%% file: content/1_introduction.tex
\section{Introduction}
\label{sec:introduction}

Allowing a user to see their body in the virtual reality (VR) world is an important part of an immersive experience. It enables natural interactions with the virtual environment and other users, and potentially allows the remote users to communicate and collaborate in a way as if they met in person. 
A VR system needs to track the full-body movement of a user in real-time and animate an avatar in the VR world that can faithfully reproduce the user's actions to support such an experience. 
However, typical consumer VR systems, such as HTC Vive and Occulus Quest, only provides up to three tracker devices in their default configurations, including a head-mounted display (HMD) and optionally two hand-held controllers (HHC). 
Estimating full-body motions with high degrees of freedom using such limited information is an ill-formed problem. While it is relatively easy to compute good-quality upper body movement with carefully tuned IK solvers when the user is sitting or standing in place, finding correct heuristics for predicting plausible footsteps and lower-body motion is usually very difficult when the user needs to move around, and the solutions are often prone to artifacts such as foot-skating and unrealistic movement.
%
%

Data-driven methods, especially those learning generative models from massive motion capture data, have been proven to be an effective way to generate high-quality interactive character animation in recent years. The latent structure embedded in the motion data helps regularize the animation process and ensure the naturalness of the generated motions. 
Physics-based methods, on the other hand, have long been a promising avenue for creating realistic character animation, where physics-based simulations naturally prevent artifacts like foot-skating and can generate physically accurate motions. The recent advances in deep reinforcement learning have demonstrated promising results that flexible motion control strategies can be learned by imitating reference motion, making the combination of the data-driven and physics-based method a possible way to address the full-body motion tracking problem in VR as discussed above.

In this work, we present a data-driven physics-based method for estimating realistic full-body movement using up to three built-in VR tracker devices included in a typical VR system.
Our system simulates an avatar character that can mimic user actions in the virtual world in real-time, which has the potential to enhance the immersive experience in VR applications. We train our system using reinforcement learning with carefully designed pretraining processes to ensure the success of the training and the quality of the simulation. 
{The principal contributions of this work include: (1) We build a novel real-time full-body motion tracking system that generate physically realistic motions from sparse VR trackers. Our configuration is directly compatible with commercial VR systems, and potentially supports HMD-only VR systems. (2) We develop a full-body motion predictor module with decoupled upper-body and lower-body pose predictors and combine them via an aggregated representation of the state of the character. We find this network architecture performs better than the baseline methods that directly predict the full-body movement and is more robust with respect to unseen upper-body motions. 
}

%% file: content/2_related_work.tex
\section{Related Work}
\label{sec:related_work}

Human motion capture plays an important role in character animation. In game and film industries, commercial motion capture solutions, such as Vicon~\cite{Vicon} and Xsens~\cite{Xsensa} has been widely adopted to capture high quality human performance. These high-end mocap systems often use tens of optical markers or IMU sensors to achieve accurate capturing and high motion quality. Such a configuration is often expensive and not suitable for a commercial VR systems in everyday settings. 
Reconstructing full-body human motions using a small number of sensors or markers becomes more and more demanding given the popularization of the VR/AR devices. Many state-of-the-art approaches achieve this goal using six body-worn sensors on the user's head, limbs, and waist \cite{Liu2006_Human,Liu2011,vonMarcard2017Sparse,Huang2018Deep}, which does not directly work with the out-of-box components of a typical VR systems. 
{Research on estimating full-body human motion using even fewer tracking signals remains relatively sparse, where a sensor mounted on waist is often needed to  reconstruct flexible lower-body motions. For example, \cite{Kim2012_Realtime,Kim2013_Human} demonstrate systems based on kernel canonical correlation analysis (CCA) to predict full-body poses from five motion sensors mounted on a user's limbs and the waist. \cite{Wouda2019Time} consider a similar setting, but employ a LSTM-based model to reconstruct the full-body motion. \cite{Yang2021LoBSTr} show that plausible lower-body motions can be generated by a GRU-based model from four upper-body VR sensors mount on head, hands, and the waist. 
DeepMotion~\cite{DeepMotion} offers a physics-based three-point full-body tracking solution. Their technique is based on a simplified control model and the simulated avatar can look robotic. 
\cite{Dittadi2021FullBody} recently shows a VAE-based model can be trained to predict full-body poses from a single head-mounted device, but their approach does not predict root transformations.}
{Unlike those approaches, our method enables natural and physically plausible full-body tracking using up to three VR trackers, which is compatible with typical commercial VR systems.
}

Synthesizing interactive human motions using low-dimensional control signals, such as those using a keyboard or a game controller to control a game character \cite{Starke2020Local,SimonClavet2015Nucl} or manipulating a virtual puppet using hand or full-body gestures \cite{Seol2013_Creature,Rhodin2014_Interactive}, has been a long-term topic of character animation. The motion graphs and its variations \cite{Kovar2002Motion,Heck2007} has been widely adopted in game industry as a standard technique, where interaction can be achieved using either hand-crafted state-machines or control policies trained using reinforcement learning~\cite{Wampler2010Character,Treuille2007}. 
Motion synthesis using unorganized motion data had attracted a lot of attentions in the past years. \cite{Lee2010Motion} introduced a framework that learns to blend the nearest neighbors of the character's pose in a dataset to achieve fast transitions between motions under user control. This method later inspired the development of Motion Matching methods~\cite{SimonClavet2015Nucl,Holden2020}.
Learning generative models from massive unorganized data is another promising way to achieve interactive control. Previous research has explored many statistic models, such as PCA~\cite{Safonova2004}, mixture of Gaussian \cite{Min2009,Min2012Motion}, and Gaussian Process \cite{Levine2012Continuous,Wang2008Gaussian}. More recently, deep generative models have demonstrated great potential in achieving realistic and interactive motion generation. Recent research has exploited popular models, such as GAN~\cite{Wang2021Combining}, autoencoder \cite{Holden2016Deep}, VAE~\cite{Ling2020Character}, and normalizing flows \cite{Henter2020MoGlow}, with many different network structures, such as CNN~\cite{Holden2016Deep}, LSTM~\cite{Lee2019Interactive,Harvey2020}, mixture of expert~\cite{Starke2020Local,Zhang2018Modeadaptive}, and transformers \cite{Li2021Learn}.
{
Our system develops a data-driven generative model to predict a full-body poses from VR tracker inputs, with a carefully designed structure that reduces the coupling between motion generations of different parts of the character, thus allowing additional robustness with respect to unseen input.
}

\begin{figure*}[t]
    \centering
    \includegraphics[width=0.9\linewidth]{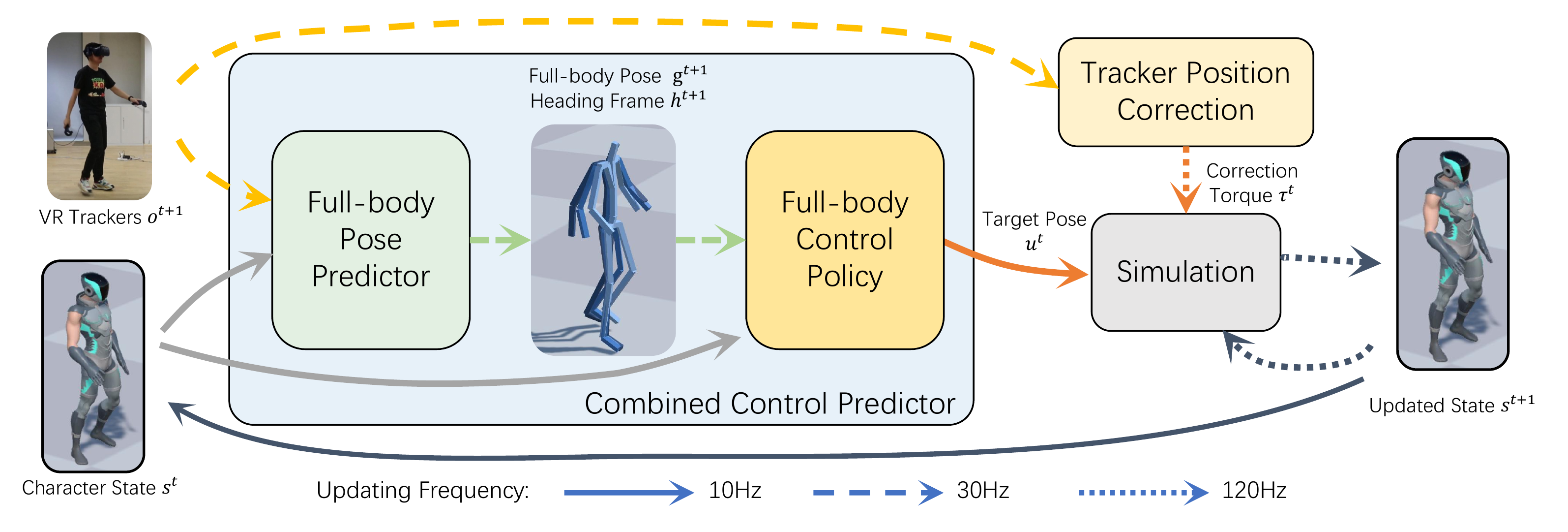}
    \caption{The architecture of our system. Our system is composed of a full-body pose predictor, a full-body control policy, and a simulation module. The pose predictor takes as input the transformations of the VR trackers and predicts the full-body pose as well as the location and heading direction of the user. The control policy uses this information to calculate target poses for the PD controllers, {combined with a tracker position correction module that compute necessary joint torques to compensate for the tracking error of the trackers}. The simulation module then performs physics simulation and updates the character's state. This new state will be used as the input of the next frame.}\label{fig:system_overview}
\end{figure*}

Unlike data-driven methods, physics-based approaches explicitly incorporate physics simulation into the motion generation pipeline, which ensures physical accuracy of the generated motions and allows responses to unexpected perturbations. 
However, designing a physics-based controller for complex human skills has been a notorious challenge in computer graphics and robotics. Early approaches often rely on hand-crafted controllers~\cite{Hodgins1995,Yin2007SIMBICON,Lee2010Datadriven}, optimized feedback policies~\cite{Tan2014Learning,Liu2012Terrain,Wang2012Optimizing}, abstract models~\cite{Coros2010Generalized,Mordatch2010Robust,Kwon2017MomentumMapped}, optimal control~\cite{Muico2009}, and model predictive controllers~\cite{Eom2019,Hong2019Physicsbased}. 
The recent advancement of reinforcement learning (RL) makes imitating motion data a feasible way to learn control policies for complex skills~\cite{Liu2016Guided,Peng2018DeepMimic,Chentanez2018}. To create a multi-skilled character, individual controllers can be organized and scheduled by high-level policies~\cite{Peng2017DeepLoco,Liu2017Learning,Merel2019Hierarchical} or be used to train an integrated policy ~\cite{Merel2019Neural,Merel2020Catch,Won2020}. Direct training of multi-skilled policies can also be achieved using a mixture of expert structure~\cite{Peng2019MCP,Luo2020CARL} or with the help of adversarial losses~\cite{Merel2017Learning}.
Combining the advantage of both data-driven motion generators and RL-based control policy is another avenue to creating multi-skilled and interactive control policies~\cite{Park2019Learning,Bergamin2019}, and similar ideas are also adopted to reduce the ambiguity caused by incomplete input signals~\cite{Ri2021Jehee,Shimada2020,Xie2021_PhysicsBaseda}.
{Our work also takes benefit of both the data-driven approaches and physics simulation to achieve physically realistic motion generation. 
}

%% file: content/3_system_overview.tex

\section{System Overview}\label{sec:system_overview}

The goal of our system is to reconstruct realistic full-body movement of the user according to the positions and orientations of three \emph{VR trackers}, \ie{} the head-mounted display (HMD) and two hand-held controllers (HHC), and to simulate an avatar character in the virtual world to reproduce the user's motion at real-time.
%
As sketched in \fig\ref{fig:system_overview}, our system is composed of {four major components} operating at different timescales. 


Our system obtains a stream of transformations of the VR trackers using the built-in functions provided by the VR systems. \lb{The frame rate of these input signals is assumed to be 30~Hz, where resampling is applied when necessary}. The \emph{Full-body Pose Predictor} module reads the transformation signals of the VR trackers and predicts the full-body pose as well as the location and heading direction of the user according to the current state of the avatar character. \lb{These estimations can then be used to update the character's pose directly. We refer to this updating strategy as the \emph{direct mode} of our system.}

\lb{The {direct mode}, however, often generates physically implausible results such as unrealistic foot sliding due to the lack of physical  constraints. To create a physics-enhanced user experience, we employ a \emph{Full-body Control Policy} module to take the estimated poses as a reference and compute a target pose that will be used to actuate the character to track the user's action. Then, the \emph{Simulation} module is involved to simulate the character. 
We refer to this process as the \emph{normal mode} of our system.}

\lb{The user often changes their movement unpredictably, making it hard to track their global position and pose accurately in simulation. For example, the avatar character may generate an excessive speed to match the user's current position without knowing that the user is intending to stop immediately, in which case the momentum of the character may prevent it from stopping quickly and thus cause tracking errors. To deal with this problem, we opt for allowing a small amount of delay in the tracking to enable the system to prepare for the unpredictable changes. Specifically, we let the {Full-body Control Policy} module to operate at a coarse timescale of 10~Hz. When every three frames of the tracker input are received, the Full-body Pose Predictor is involved to predict a series of poses from the current simulation states of the character in an autoregressive manner. These estimations are then used by the {Full-body Control Policy} to compute the target pose. We thus refer to the full-body pose predictor and control policy jointly as a \emph{Combined Control Predictor}.}


The \emph{Simulation} module performs physics simulation and updates the state of the character. 
In our system, the character is modeled as an articulated rigid body skeleton with a floating root, where PD-servos are employed to track the target poses provided by the control predictor and compute joint torques to actuate the internal degrees of freedom of the character. \lb{We run simulation at a relatively high frequency, 120~Hz, to ensure numerical stability. The same target pose is used at every simulation step until the control predictor computes a new target. } 

Finally, an additional \emph{Tracker Position Correction} module is involved to encourage the simulated character to follow the VR trackers accurately. This module operates at the same timescale as the simulation. {It applies virtual forces to the hands of the simulated character, where the force is computed with a PD controller according to the tracking errors.
These virtual forces are implemented as additional joint torques computed using Jacobian transpose control and applied to the corresponding arms.}


The system is trained using the deep reinforcement learning framework by imitating prerecorded motions. To achieve good performance, we have captured {one hour of unorganized performance data, where the subjects were asked to stand or walk in the capture volume while acting as if they were playing a VR game.} We find a vanilla end-to-end training hard to converge for such an integrated system with many coupled components. To facilitate the training, we pretrain the full-body pose predictor using supervised learning and the control predictor by learning to track the reference motions, and then fine-tune them jointly using reinforcement learning.



%% file: content/4_control_predictor.tex
\section{Pose Predictor}\label{Sec:pos}

The \emph{full-body pose predictor} module of our system, represented by $\mathcal{G}$, estimates the user's pose and heading transformation according to the VR tracker input and the past states of the avatar character.
Formally, the input to this module is a stream of transformations of the three VR trackers $\vrc = \{\bm{p}_{X}, \bm{q}_{X}\}$, where $\bm{p}_X$ and $\bm{q}_X$ are the location and orientation of tracker $X$ in the world coordinate frame, and $X\in\{\text{H}, \text{L}, \text{R}\}$ corresponds to the HMD and left/right hand-held controllers, respectively.
Unless otherwise noted, {we use quaternions to represent the orientation of the VR trackers and 3-D rotation vectors for the joint rotations and bone orientations of the character in this paper}.



\begin{figure}[t]
    \centering
    \includegraphics[width=\linewidth]{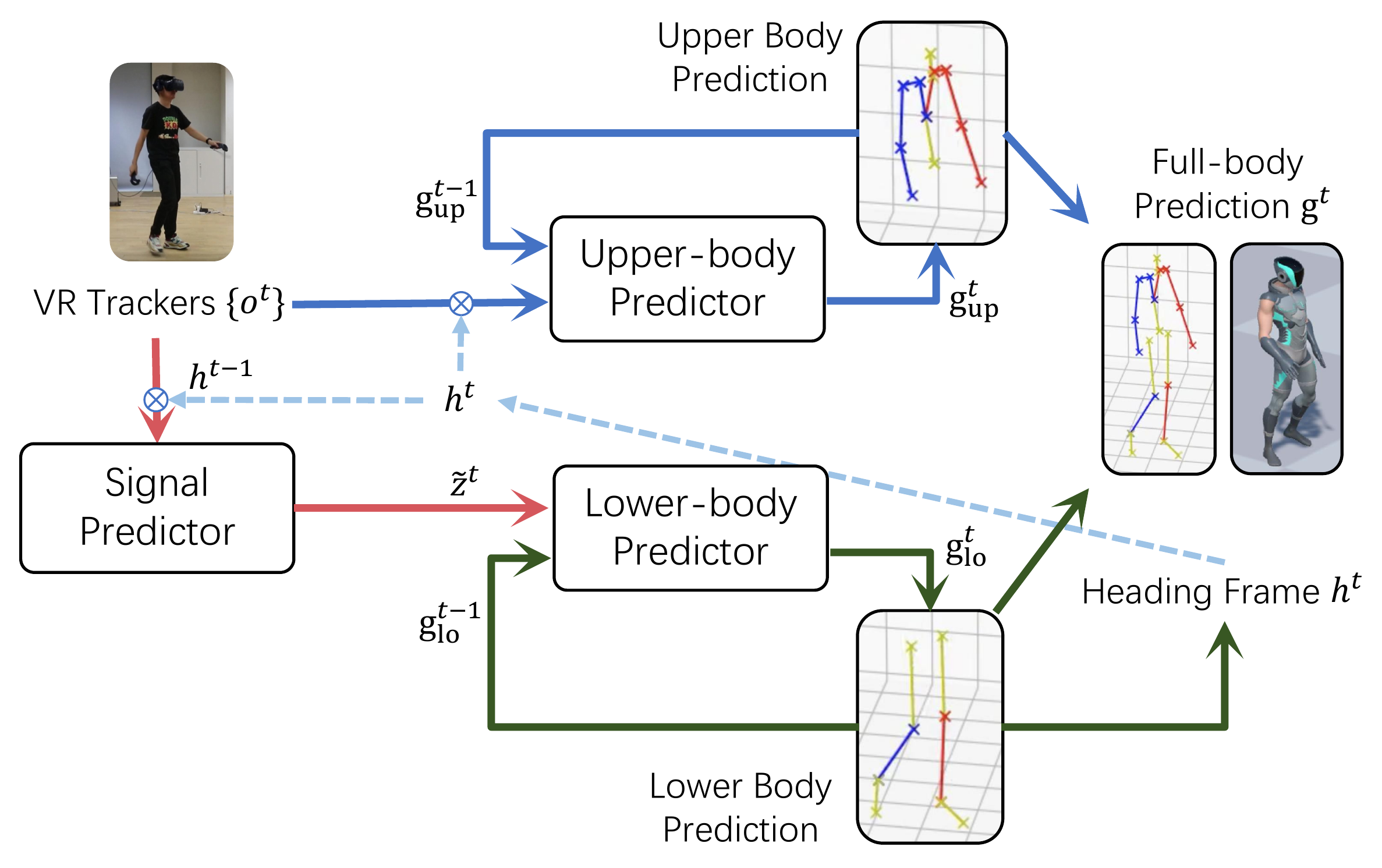}
    \caption{{The architecture of our pose predictor}}
    \label{fig:pose_predictor}
\end{figure}

We formulate the pose predictor $\mathcal{G}$ as a recurrent model. Given a sequence of VR tracker input $\{\vrc^t\}, t=1,\dots,T$, an initial state of the character $\genstate^0$, and an initial heading transformation $\heading^0$, the pose predictor generates a series of states $\{\genstate^t, \heading^t\}, t=1,\dots,T$ auto-regressively as
\begin{equation}\label{eqn:pose_predictor}
    (\vrc_*^{t+1},\genstate^{t},\heading^{t}) = \mathcal{G}(\vrc^t, \genstate^{t-1}, \heading^{t-1}),
\end{equation}
where the heading transformation $\heading$ horizontally moves with the root of the character and has one axis vertically aligned and another aligned with the character's heading direction. 
We use a 3-tuple $(p_x, p_z, \theta_y)$ to represent $\heading$, where $(p_x, p_z)$ is the planar translation of the coordinate frame and $\theta_y$ corresponds to the rotation around the vertical axis. 
{Inspired by \cite{Lee2019Interactive}, we let $\mathcal{G}$ also predict the VR tracker input of the next time step, $\vrc_*^{t+1}$.} 

When using a VR application, a user's upper body typically exhibits higher range of motion than his lower body, and  the movements of the upper body and the lower body are not always {strongly correlated}. For example, a user can swing his arms in multiple ways while either standing in-place or walking around. 
To deal with such {degrees of freedom}, we consider our character as two disjoint sets of joints and treat them separately. As shown in \fig\ref{fig:pose_predictor}, a lower-body pose predictor, $\mathcal{G}_\lob$, controls the joints of the character's legs, represented by $J_{\lob}$, 
while an upper-body pose predictor $\mathcal{G}_\upb$ handles the set of joints of the upper body and arms of the character, $J_{\upb}$. 
{More specifically, given the locations and orientations of the three VR trackers, the lower-body pose predictor $\mathcal{G}_\lob$ predicts the global motion of the user in terms of the movement of the heading frame $\heading$, and computes coordinated leg motions. The upper-body pose predictor $\mathcal{G}_\upb$ then computes an upper-body pose that follows the trackers' position in the predicted heading frame}.

The state of the character is then represented as $\genstate=\{\genstate_\upb, \genstate_\lob\}$, where $\genstate_\upb=\{\bm{p}_j, \bm{v}_j, \bm{q}_j\},j\in{}J_{\upb}$ and $\genstate_\lob=\{\bm{p}_j, \bm{v}_j, \bm{q}_j, \contact, \genctrl\},j\in{}J_{\lob}$ 
consist of the position $\bm{p}_j$, velocities $\bm{v}_j$, and orientations $\bm{q}_j$ of the joints in the corresponding joint sets $J_{\upb/\lob}$, all computed in the reference heading coordinate frame $\heading$. 
{$\contact=\{c_{\text{L}}, c_{\text{R}}\}$ contains two scalar variables $c_{\text{L/R}}\in[0,1]$ indicating if the left foot and right foot are in contact with the ground or not, respectively}.
{To lower the coupling between the two pose predictors $\mathcal{G}_\lob$ and $\mathcal{G}_\upb$}, we utilize an aggregated state of motion of the upper body, $\genctrl$, to convey necessary information to the lower body,
where $\genctrl=\{\bm{p}_{\upb}, \bm{v}_{\upb}, \bm{L}_{\upb}, \delta\heading\}$ consists of the centroid position $\bm{p}_{\upb}$, velocity $\bm{v}_{\upb}$, and angular momentum $\bm{L}_{\upb}$ of the character's upper body, as well as  the change of heading frame $\delta\heading$ from the last time step.
%
%
All these quantities are computed in the current reference heading coordinate frame $\heading$, except for $\delta\heading$, which is computed with respect to the heading frame of the previous time step, so that the heading frame can be updated as $\heading^t = \heading^{t-1} \otimes \delta\heading^{t}$, where $\otimes$ represents the multiplication of two transformations.
{At runtime, a dedicated \emph{signal predictor} $\mathcal{G}_\hb$ is employed to predict this aggregated state $\genctrl$ from the input signals of the three VR trackers~$\vrc$, allowing the lower-body pose predictor to focus on the global motion of the user and less distracted by the diversity of the upper body motions.}

As sketched in \fig\ref{fig:pose_predictor}, the entire pose prediction process of \eqn\eqref{eqn:pose_predictor} can then be rewritten as
\begin{align}\label{eqn:pose_predictor_detail}
    %
    \tilde{\genctrl}^t &=\mathcal{G}_{\hb}\left(\text{inv}( \heading^{t-1}) \otimes \vrc^t \right) \nonumber\\\
    (\genctrl^{t+1}_*, \genstate_{\lob}^t) &= \mathcal{G}_{\lob}(\tilde{\genctrl}^t, \genstate_{\lob}^{t-1}) \nonumber\\\
    \heading^t &= \heading^{t-1}\otimes{}\delta\heading^t, \quad{}\text{where\ }\delta\heading^t \in \genstate_\lob^t \\
    %
    %
    ({\bm{o}}^{t+1}_*, \genstate_{\upb}^t) &= \mathcal{G}_{\upb}\left(\text{inv}(\heading^{t}) \otimes \vrc^t, \genstate_{\upb}^{t-1}\right) \nonumber\\\
    \genstate^t &\leftarrow \{ \genstate_\upb^t, \genstate_\lob^t \}  \nonumber\,
\end{align}
where the operator $\text{inv}$ and $\otimes$ represent the inversion and multiplication of transformations, respectively.
Specifically, when receiving a set of VR tracker signals $\vrc^t$ at time $t$, the signal predictor $\mathcal{G}_{\hb}$ first transforms $\vrc^t$ into the current heading frame ${\heading}^{t-1}$ and computes a predicted aggregated upper-body state $\tilde{\genctrl}^t$ accordingly. 
The lower-body predictor $\mathcal{G}_{\lob}$ then takes $\tilde{\genctrl}^t$ as input and predicts the lower-body state $\genstate_\lob^t$.
%
%
After updating the heading frame using the predicted $\delta\heading^t$, which is included in $\tilde{\genctrl}^t$, the VR tracker input $\vrc^t$ is transformed into the new reference heading coordinate frame ${\heading}^t$, and then the upper-body predictor $\mathcal{G}_{\upb}$ estimates a new upper-body pose $\genstate_{\upb}^{t}$ according to it. 
{Note that both $\mathcal{G}_\lob$ and $\mathcal{G}_\upb$ also predict $\genctrl_*^{t+1}$ and $\vrc_*^{t+1}$, the corresponding input signals of the next time step respectively.} 
Finally, the predicted upper-body and lower-body states $\genstate_{\upb}^{t}$ and $\genstate_{\lob}^{t}$ are put together to construct the full-body state $\genstate^t$. 
When a series of tracker input $\{\vrc^t\}$ is given, the pose predictor module repeats the above procedure and generates a sequence of states, forming a motion clip.

\subsection{Pose Predictor Training}
We implement the three sub-predictors: the signal predictor $\mathcal{G}_{\hb}$, the lower-body predictor $\mathcal{G}_{\lob}$, and the upper-body predictor $\mathcal{G}_{\upb}$, as recurrent neural networks, each consisting of three GRU (Gated Recurrent Unit) layers, combined with additional one-layer fully-connected encoder and decoder layers. 
The number of hidden layer units of each network is set as 64, 128 and 128, respectively. 

We employ a two-stage training process to train the pose predictor using the motion capture data, where each sub-predictor is pretrained separately with the input/output extracted from the motion data and then fine-tuned jointly while following the combined prediction process described above. 
Each training episode starts from a batch of starting states randomly chosen from the motion dataset, followed by generating a sequence of states of length $T$ in an auto-regressive manner with the corresponding inputs extracted from the motion data. The objective functions are then evaluated on each generated state, whose gradients are used to update the networks. 
We use $T=60$ and a batch size of $32$ during the training. 
Two extra dropout layers with dropout rates of 0.1 and 0.05 respectively are applied after the encoder layer and before the decoder layer of each sub-predictor to prevent over-fitting during training. 
\lb{We use the Adam optimizer~\cite{Kingma2014Adam} to perform the gradient update with a learning rate of $5\times 10^{-4}$.}

The loss functions for the three sub-predictors are defined as
\begin{align}
    \mathcal{L}_{\hb} &= \mathcal{L}^{\mse}_{\tilde{\genctrl}} + w_{\heading}\mathcal{L}_{\heading}\\
    \mathcal{L}_{\lob} &= \mathcal{L}^{\mse}_{\genstate_{\lob}, \tilde{\genctrl}_*} + w_{\heading}\mathcal{L}_{\heading} + w_{\text{FK}}\mathcal{L}_{\text{FK}}\\
    \mathcal{L}_{\upb} &= \mathcal{L}^{\mse}_{\genstate_{\upb}, \tilde{\vrc}_*} {+ w_{\text{FK}}\mathcal{L}_{\text{FK}}},
\end{align}
respectively, where the loss terms in the form of $\mathcal{L}^{\mse}_{*}$ are weighted MSE (mean squared error) between the quantities represented by the {subscripts} and their corresponding ground truth. 
When training the signal predictor $\mathcal{G}_{\hb}$ and the lower-body predictor $\mathcal{G}_{\lob}$, an additional loss term $\mathcal{L}_{\heading}$ is used to minimize the error of the estimated global heading transformation, computed by accumulating the sequence of predicted change of heading frames. The global heading transformation can change dramatically in a long motion sequence. To avoid the singularity of the rotation angle representation, we compute $\mathcal{L}_{\heading}$ as
\begin{align}
    \mathcal{L}_{\heading} &= \mse_t\left[(p_x^t, p_z^t), (\bar{p}_x^t, \bar{p}_z^t)\right] \nonumber\\
    &+
    \mse_t\left[(\sin\theta_y^t, \cos\theta_y^t), (\sin\bar{\theta}_y^t, \cos\bar{\theta}_y^t)\right],
\end{align}
where the symbols with a bar $(\bar{\;})$ indicates ground truth variables. 
In addition, to ensure consistency between the predicted joint orientations and joint positions, we perform forward kinematics (FK) according to the predicted joint orientations $\{\bm{q}_j\}, j\in{}J_{\upb/\lob}$ and compute the MSE between the resulting joint positions and the predicted ones $\{\bm{p}_j\}, j\in{}J_{\upb/\lob}$. We consider this MSE as an extra loss term $\mathcal{L}_{\text{FK}}$ in the training.

In the fine-tuning process, we train the entire pose predictor while minimizing the combined loss $\mathcal{L}=\mathcal{L}_{\hb}+\mathcal{L}_{\lob} +\mathcal{L}_{\upb}$. {In practice, we  freeze the parameters of the upper body predictor $\mathcal{G}_{\upb}$ and only update the signal predictor $\mathcal{G}_\hb$ and the lower body predictor $\mathcal{G}_\lob$ in this process, which helps stabilize the training and prevent degeneration.} 

\section{Full-body Control Policy}

As shown in the system overview of \fig\ref{fig:system_overview}, the \emph{Full-body Control Policy} module, represented by $\pi$,
converts the estimated full-body poses of the user into target poses, which will be used to actuate the simulated character using PD-servos. 
%
More specifically, the input to $\pi$ is a state vector $\rlstate^t=\{\genx^t, \genstate^{t+k}\}, k=1,2,3$, which contains the current simulation state $\genx^t = \{\bm{p}_j^t, \bm{v}_j^t, \bm{q}_j^t\},j\in{}J$ and three predicted states $\{\genstate^{t+k}\}, k=1,2,3$ corresponding to $0.1$, $0.2$, and $0.3$ seconds after the current time $t$ respectively. 
%
We convert $\genx^t$ into a reference global heading frame to inform the simulated character about the global tracking errors, which helps the character track the heading of the user accurately.  $\{\genstate^{t+k}\}$ are expressed in the same heading frame as well. 
In the pretraining process, the reference heading is extracted from the reference motion. {During the finetuning and at runtime, the reference heading is comuted by the Full-body Pose Predictor.}
The output of the policy is a target pose $\rlaction=\{\bm{q}_j\}$ consisting of target rotations $\bm{q}_j$ of every internal joint $j\in{}J$.
We implement the policy $\pi$ as a simple feedforward neural network consists of four fully connected layers, with 256 units in each of the two hidden layers and ReLU as the activation function.

\subsection{Full-body Control Policy Pretraining}
We pretrain the full-body control policy using reinforcement learning on our motion capture dataset. Following the standard formulation of a reinforcement learning problem, the training process maximizes the expected cumulative reward
\begin{equation}
    \pi^* = \argmax_{\pi} \mathbb{E}_{\tau\sim\pi} \left[ \sum_t \gamma^t R(\rlstate^t) \right]
\end{equation}
over all simulation trajectories $\tau=\{\rlstate_0,\rlaction_0,\rlstate_1,\rlaction_1,\dots{}\}$ induced by $\pi$,
where $\rlstate\in{}\rlState$ is the state vector, $\rlaction\in{}\rlAction$ is the action vector that stacks the joint rotations of a target pose, $R(\rlstate^t)$ is the reward of state $\rlstate^t$, and $\gamma$ is the discount factor, which is set to $0.99$ in our system.

We train the control policy using the PPO algorithm~\cite{Schulman2017Proximal}, which alternates the collection of simulation rollouts and the policy update. \lb{During training, each simulation rollout is initialized using a random state extracted from the reference motion and ends when either it is 300 control steps long or the character falls. {A batch of 4000 control steps is collected in every update iteration}. A separate value network $V$ is used to compute values for each state, which is a feedforward network with two hidden layers of 256 units each. We use the Adam optimizer~\cite{Kingma2014Adam} to perform the gradient update. {The learning rates for the policy network and the value network are $1\times10^{-5}$and $1\times10^{-4}$, respectively.}}

\subsubsection{Reward}
The goal of the pretraining is to make our simulated character to track the reference motions accurately, thus the reward at each time step is defined as
\begin{align}
	R(\rlstate) = \exp(&w_{\text{pose}}r_{\text{pose}} + w_{\text{orient}}r_{\text{orient}} + w_{\text{pos}}r_{\text{pos}} + \nonumber\\
    &w_{\text{balance}}r_{\text{balance}} + {w_{\text{contact}}r_{\text{contact}}} +w_{\text{foot}}r_{\text{foot}}).
    \label{eqn:rl_reward}
\end{align}
%
The pose reward $r_{\text{pose}}$ encourages the character to match the reference pose locally, which is computed as
\begin{equation}
	r_{\text{pose}} = -\frac{1}{|J|}\sum_{j\in{}J} \Vert \bar{\bm{q}}_j\bm{q}_j^{-1} \Vert_A + 0.1\Vert\bar{\bm{\omega}}_j - \bm{\omega}_j \Vert,
\end{equation}
where $\Vert \bm{q} \Vert_A$ computes the angle of rotation of a quaternion $\bm{q}$, 
and $\bm{\omega}_j$ is the joint's rotational speed. The symbols with a bar ( $\bar{\;}$ ) indicates the reference.

The link orientation and position reward penalizes the global tracking errors, in terms of the positions and orientations of the character's bones:
\begin{align}
	r_{\text{orient}} &= -\frac{1}{|J|}\sum_{j\in{}J} \Vert \bar{\bm{q}}_j \ominus \bm{q}_j \Vert \\
    r_{\text{pos}} & = -\frac{1}{|J|}\sum_{j\in{}J} \Vert\bar{\bm{p}}_j - \bm{p}_j \Vert,
\end{align}
where all the quantities are compared in the global coordinate frame. 

The balance reward encourages the relative position between the character's feet and its center of mass to match those from the reference, which helps the character stabilize its pose and maintain balance. We compute this term as 
\begin{equation}
	r_{\text{balance}} = -\sum_{X\in\{\text{L,R}\}} w_X (\Vert \bar{\bm{d}}_X - \bm{d}_X\Vert + \Vert \bar{\bm{v}}_X - \bm{v}_X\Vert),
\end{equation}
where $\bm{d}_X=\bm{c}-\bm{p}_X$ is the distance between the center of mass $\bm{c}$ and the corresponding foot, and $\bm{v}_X$ is the velocity of the foot. We assume that a standing foot contributes more to balance than a moving foot, thus the weight of each reward term is adjusted according to the velocity of the feet as
\begin{align}
    w_{X} &= \frac{\hat{w}_{X}}{\hat{w}_R+\hat{w}_L} \\
    \hat{w}_{X} &= \Vert \bm{v}^{\shortparallel}_{X} \Vert + 5\max(0, h_X-\epsilon),
\end{align}
where $\bm{v}^{\shortparallel}_{X}$ represents the planar components of $\bm{v}_X$, and $h_{X}$ is the height of the foot.

{The contact reward penalizes the difference between the position of the character's foot and its reference when the foot is in contact with the ground:
\begin{equation}
	r_{\text{contact}} = -\sum_{X\in\{\text{L,R}\}} c_X\Vert \bar{p}_X - p_X\Vert,
\end{equation}
where $c_X$ is the contact label of the foot predicted by the pose predictor.}

At last, the foot height reward encourage the character to lift its feet to clear the ground when moving, which is computed as
\begin{equation}
	r_{\text{foot}} = -\sum_{X\in\{\text{L,R}\}} \Vert \bar{h}_X - h_X\Vert.
\end{equation}

\lb{In training, all of the weights in \eqn\eqref{eqn:rl_reward} are set as 10 except that $w_{\text{pose}}$ is set as 5.}

\subsubsection{Adaptive State Initialization} As also observed by previous works~\cite{Park2019Learning,Chentanez2018}, when training a tracking policy to perform a diverse set of motions, choosing the random initial states uniformly can cause the training to overfit to the motions that are easy to learn. To mitigate this problem, we employ an adaptive state initialization strategy which encourages the training process to start a rollout from a less visited state. 
Specifically, we draw initial states from a multinomial distribution, with the probability periodically updated so that a state is chosen based on its value. A state with lower value will have higher chance to be selected as a starting state. Specifically, the probability is updated by 
\begin{equation}
    P(\bar{\rlstate}) = \frac{\exp(-V(\bar{\rlstate})/T_w)}{\sum_{\bar{\rlstate}} \exp(-V(\bar{\rlstate})/T_w)},
\end{equation}
where $V$ is the value function of the RL problem. The temperature parameter $T_w$ is empirically set to 5. {During the training, we update this distribution of the initial states every 100 training iterations.}

\subsubsection{Pre-Pretraining}
{We jump-start the training by initializing the tracking policy $\pi$ using supervised learning, which significantly accelerate the training at the early stage. The training data of this pre-pretraining is created by constructing open-loop control trajectory for our motion capture data using the SAMCON algorithm~\cite{Liu2010Samplingbased,Liu2015Improving} and extracting corresponding state-action pairs from the simulation. To prevent overfitting, we apply dropout before each fully connected layers of $\pi$ with the dropout rate of (0.1, 0.1, 0.05, 0.02) respectively. These dropout layers are disabled in the above reinforcement learning process.}

\section{Combined Control Predictor}\label{sec:CombinedControl}

\yyj{As depicted in \fig\ref{fig:system_overview}, the two core components of our system, the Full-body Pose Predictor $\mathcal{G}$ and the Control Policy $\pi$, run jointly as a \emph{Combined Control Predictor.} To prepare for unpredictable movement from users, a little delay in the tracking is allowed. Specifically, the Full-body Pose Predictor module operates at a coarse timescale of 10~Hz. When every three frames of the tracker input are received, a \emph{signal forecasting} strategy is involved to predict six future frames of the input as: 
\begin{align}    
    \vrc^t &= \delta\vrc + \vrc^{t-1} \quad{} t>3
\end{align}
where $\delta\vrc$ is the average offset between consecutive input frames. The Full-body Pose Predictor then takes these extended tracker inputs and predicts a short motion clip of nine frames recursively. The initial pose of this prediction is extracted from the current state of the simulated character, {while the previous prediction of the user's heading is employed as the the initial heading}. 
{We further apply inverse kinematics in the way similar to \cite{Zhang2018Modeadaptive} to ensure accurate tracking of the input trackers.}
The Control Policy module then compute a target pose according to this reference motion clip. 
At last, the simulation advances at 120~Hz until the next 0.1-second interval starts.
}

\subsection{Combined Control Fine-tuning}
After pretraining the full-body pose predictor and the control policy, our system fine-tunes them together as the combined control predictor using an additional reinforcement learning process. 
During the training, we extract transformations of the VR trackers from the motion capture data, and encourages the character to match the reference motions as closely as possible. The same rewards as the pretraining of the full-body control policy are used in this fine-tuning process.
{We use PPO again to train the combined policy, and a progressive learning approach is adopted to facilitate the training. The tracking position correction and the signal forecasting components are disabled at the beginning until the training process has plateaued. Then, the training continues with all the components enabled, while the motion generated by the pose predictor is used as the reference for the rewards.}
We find that the pose predictor $\mathcal{G}$ is prone to degeneration in the training without effective regularization. To mitigate this issue, the parameters of $\mathcal{G}$ is frozen in the fine-tuning. 



\subsection{Tracker Position Correction}
\lb{To ensure accurate tracking of the VR trackers, we employ an additional Tracker Position Correction module in our system. 
This module applies virtual forces on the avatar's hands to correct their position. Each force is computed using PD control:
\begin{equation}
    F_{virtual} = k_p * (\bar{\bm{p}} - \bm{p}) - k_d * \dot{\bm{p}},
\end{equation}
where $\bm{p}$ is the position of a hand and $\dot{\bm{p}}$ is its linear velocity. $\bar{\bm{p}}$ is the position of the corresponding VR tracker. $k_p$ and $k_d$ here are set to 1000 and 10, respectively.}
\lb{These virtual forces are not directly applied to the hands of the simulated character. Instead, the character tries to realize such virtual forces using joint torques. The joint torques are computed using Jacobians transpose control, thus ensuring zero accumulated external forces and torques so that the simulation is still physically correct.}

%% file: content/5_implementation.tex
\section{Implementation Details}\label{sec:implementation}
Our system is implemented in python, where the networks are built and trained with PyTorch. We simulate an avatar character that is $1.75$\,m tall and weighs $63$\,kg using a \lb{proprietary} physics engine based on joint dynamics in generalized coordinates. The character is modeled as an articulated rigid body skeleton with a floating root and actuated by PD-servos. We employ implicit joint damping to stabilize the PD control as suggested by several previous works~\cite{Tan2011,Liu2016Guided,Peng2018DeepMimic}, which allows stable simulation with a relatively large time step at 120\,Hz. The PD-gains of the PD-servos are set to $k_p=200$ and $k_d=20$ for all the joints of the character.

\lb{We have motion captured one hour of unorganized performance data using an OptiTrack~\cite{OptiTrack} motion capture system. The subjects were asked to stand or walk in the capture volume while acting as if they were playing a VR game. The motion are then retargeted to the simulated avatar by copying the rotations of the corresponding joints.}
{We implement our system on a computer with Intel Xeon Gold 6252 CPU (24 cores, 2.10~GHz). The pre-training of the pose predictor and the tracking policy takes about 72 hours. And then fine-tuning the pose predictor takes about 48 hours. Furthermore, it requires about 120 hours for combined control fine-tuning. In conclusion, the overall training procedure needs 240 hours. }

The VR environments are built with Unity and {SteamVR} plugin to communicate with a HTC Vive VR system. The VR applications and our framework in python run in separate processes on a modern computer with a multicore CPU and communicate with each other through a TCP connection. The entire system runs faster than real time, ensuring a smooth user experience without lagging.


To allow our framework to work with different VR systems with HMD and hand-held controllers of various sizes, {we consider the input to the full-body pose predictor as the location of the user's wrists and neck.} {A calibration processes is implemented to convert the true transformations of the VR trackers into the corresponding input signals, which is achieved by asking the user to perform a T-pose and measuring the distance between the VR devices and the corresponding joints.}

The pose predictor $\mathcal{G}$ can be used as a standalone module where the output full-body state is used to drive the avatar character directly, which we refer to as the \emph{direct mode} of our system, {as oppose to the \emph{normal} model where the control and simulation are involved}. We adopt an extra inverse kinematics approach similar to that was used in {\cite{Starke2020Local}} to enforce foot contacts and achieve accurate tracker positions in this mode. 


       


%% file: content/6_results.tex
\section{Results}
\begin{figure*}[t]
    \begin{subfigure}[b]{0.49\linewidth}
    \centering
    \includegraphics[width=\linewidth]{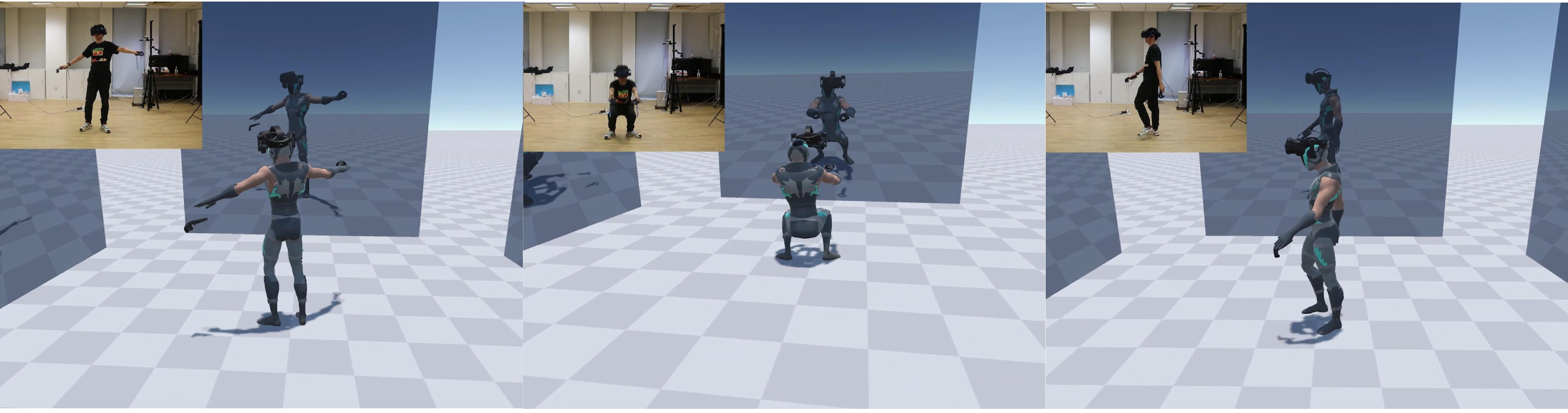}
    \vspace{-5pt}
    \caption{}\label{fig:res:mirror}
\end{subfigure}
\begin{subfigure}[b]{0.49\linewidth}
    \centering
    \includegraphics[width=\linewidth]{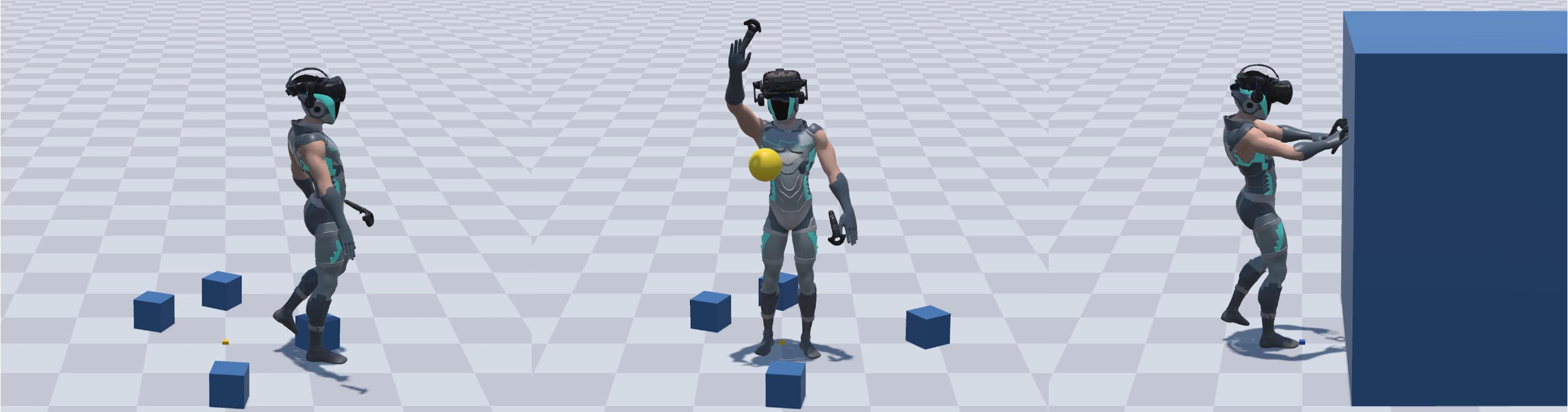}
    \vspace{-5pt}
    \caption{}\label{fig:res:interact}
\end{subfigure}

\begin{subfigure}[b]{0.49\linewidth}
    \centering
    \includegraphics[width=\linewidth]{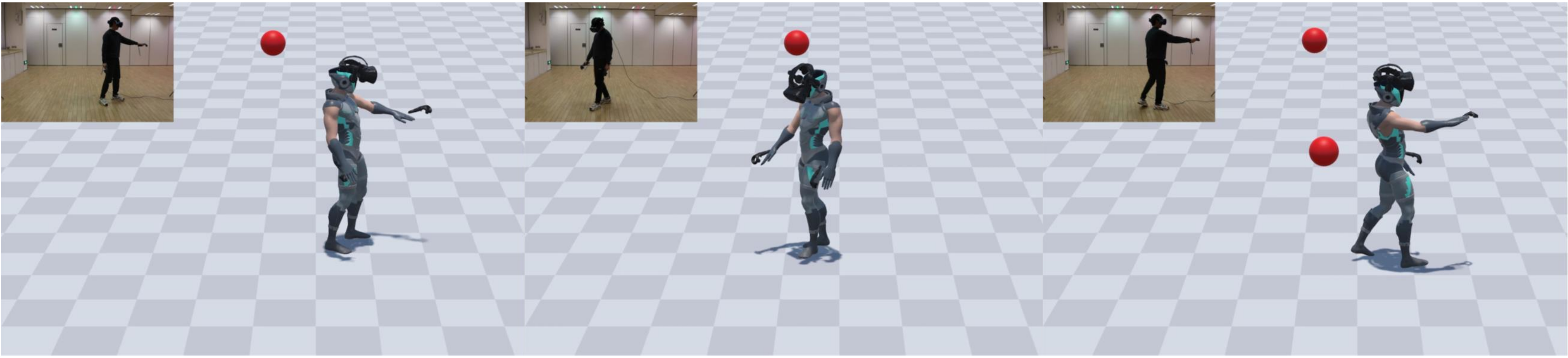}
    \vspace{-5pt}
    \caption{}\label{fig:res:ball_fetch}
\end{subfigure}
\begin{subfigure}[b]{0.49\linewidth}
    \centering
    \includegraphics[width=\linewidth]{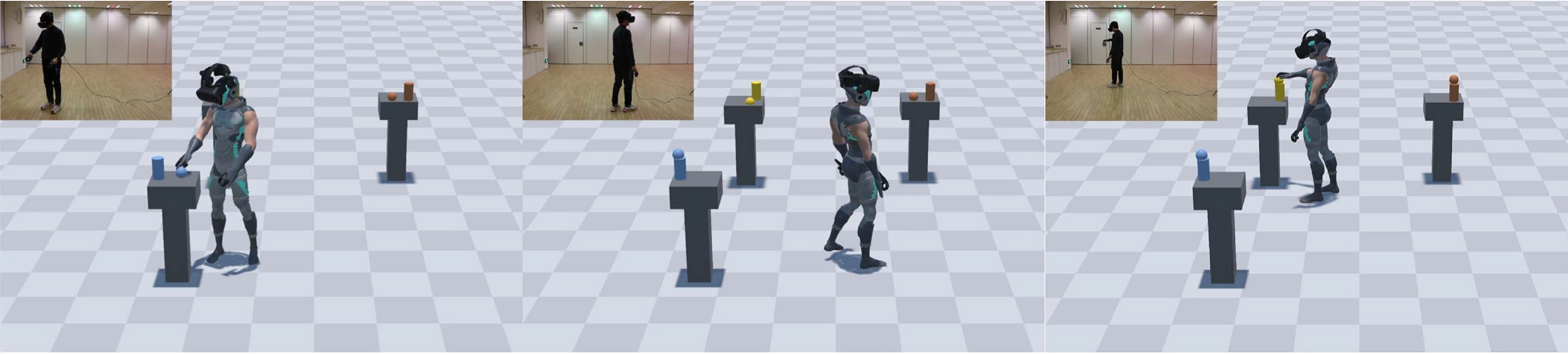}
    \vspace{-5pt}
    \caption{}\label{fig:res:fetch_place}
\end{subfigure}
\caption{Four test VR scenes. (a) A user stands, walks, and squats while swinging arms in the virtual environment. Mirrors are placed around the user to allow {them} to check the action of the virtual avatar. (b) {A participant interacts with simulated objects by stepping on or pushing them while being shot by balls from random directions}. (c) {A player walks around and eliminates balls randomly spawned in the scene}. (d) {A participant rearranges the objects with the same color onto the same tables.}}
\end{figure*}

\begin{figure}[t]
    \centering
    \includegraphics[width=\linewidth]{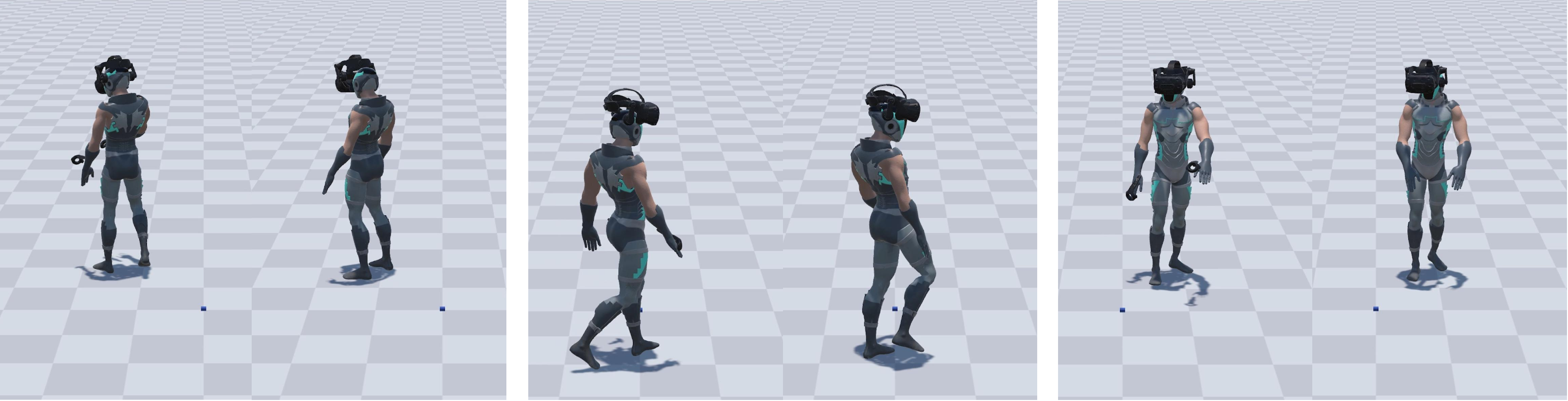}
    \caption{{Comparison between one-point tracking and three-point tracking. The character on the left of each figure uses the input of all three trackers. The character on the right uses the same set of input, but only the head tracker is considered.}}\label{fig:res:one_point_tracking}
\end{figure}

We demonstrate a variety of examples where a user plays in VR with our system. These are best seen in the accompany video. Snapshots of these examples are available in \fig\ref{fig:teaser}, {\ref{fig:res:mirror},
\ref{fig:res:interact}, \ref{fig:res:ball_fetch}, \ref{fig:res:fetch_place}} and \ref{fig:res:one_point_tracking}.
We also conduct ablation studies to validate our design of the system.

\subsection{Three-Point Tracking in VR Scenes}
We first test our full-body three-point tracking system for a wide range of motion performed by a participant in a VR environment. Before playing, the participant is required to perform a T-pose, when the distance between the VR trackers and the corresponding joints are measured as described in Section~\ref{sec:implementation}.

We have implemented four VR scenes for this test. In the first VR scene, the user can move freely in the VR environment. Several virtual mirrors are placed around the user so that {they} can check the action of his virtual avatar in real-time. {In the second scene, we put several simulated objects around the character and shoot it using balls from random directions. The user can interact with these objects by stepping on or pushing them, while {their} avatar responds to the environment changes or when hit by the balls}. The third VR scene is a mini-game, where the player needs to move and eliminate randomly spawned objects by touching them. In the last VR environment, the participant is asked to rearrange objects of different shapes and move objects with the same color together.

We test the performance of the system in both its normal model and the direct mode as described in Section~\ref{sec:implementation}. In all the experiments, our system successfully reconstruct the full-body movement of the user from the input of the three VR trackers. 
The direct mode has better tracking accuracy while suffering from unstable foot stepping and foot sliding occasionally. In contrast, the normal mode is more robust to unseen input and has no foot sliding. {The motion generated in the normal mode is more natural and has realistic physical details. For example, when the character is standing and waving his hand, the upper body in the normal mode will sway slightly with the waving hand, which can not been generated in the direct mode. However, in the normal mode, the user can experience a short delay and less tracking accuracy.} The tracking accuracy can be referred in the first-person perspective scene placed in the lower-left corner of each result in our supplementary video.

\subsection{One-Point Tracking with HMD}
We train our full-body pose predictor model to also predict the VR tracker input of the next time step, as indicated in \eqn\eqref{eqn:pose_predictor}, which allows our system to reconstruct a full-body motion with fewer or even a single VR tracker. It is often the case when a VR system provides only a head-mounted display (HMD) in its default configuration, such as the Google Daydream.
\lb{We demonstrate this capacity of our system by excluding the signals of the two hand-held controllers from the VR tracker input and only use the head tracker to reconstruct the full-body motion. Note that we have made our full-body pose predictor to predict the transformations of the trackers of the next frame in \eqn\eqref{eqn:pose_predictor_detail}. Our system thus considers these estimations as the input of the missing hand-held controllers.} 
{As shown in \fig\ref{fig:res:one_point_tracking}} and also in the supplementary video, our system successfully reconstructs plausible full-body poses using only the HMD input.

\subsection{Choice of Network architecture}
We employ a decoupled network architecture for the full-body pose predictor module $\mathcal{G}$, where an aggregated representation of the upper-body state and the global motion of the user is used to convey necessary information among sub-predictors. We believe this architecture helps reduce the coupling between each individual component and improves the robustness of the predictor with respect to unseen inputs. 

{To validate this design, we compare the performance of our network architecture with two baseline networks: a RNN-based full-body pose generator  $\mathcal{G}_{\text{full-body}}$ and a phase-functioned neural network $\mathcal{G}_{\text{PFNN}}$~\cite{Holden2017Phasefunctioned}. $\mathcal{G}_{\text{full-body}}$ is implemented to replace the decoupled networks $\mathcal{G}_\lob$ and $\mathcal{G}_\upb$. It takes the same input but computes the full-body motions directly. The network structure of $\mathcal{G}_{\text{full-body}}$ is the same as $\mathcal{G}_\lob$, except for the dimension of the input and output layers. This structure is inspired by \cite{Lee2019Interactive}, but we employ GRUs instead of the LSTM as the recurrent units.}
%
%
{The phase-functioned neural network $\mathcal{G}_{\text{PFNN}}$ is implemented similarly to \cite{Holden2017Phasefunctioned}, where four expert networks are blended using the phase parameter computed according to the contact labels. We model these experts as feed-forward neural networks consisting of three full-connected layers, and the number of hidden layer units is set to 512. The input to $\mathcal{G}_{\text{PFNN}}$ is a sequence of recorded tracker signals $\{\vrc^{t+k}\}$ converted into the current heading frame, where $t$ represent the current frame, and $k\in[-0.5\,s,0.3\,s]$ sampled every 0.1 seconds. 
}


We compare the performances of these three network architectures, $\mathcal{G}_{\text{ours}}$, $\mathcal{G}_{\text{full-body}}$, and $\mathcal{G}_{\text{PFNN}}$, in the \emph{direct mode} of our system. 
{Inspired by \cite{Starke2020Local}, the assessment is based on the contact accuracy measured as the amount of foot skating, or the average horizontal velocity of the foot when it is considered to be in contact with the ground. In this experiment, we consider both the contact label predicted by the network and that computed based on the height of the foot to calculate the amount of the foot skating. }

\begin{figure}[t]
    \centering
    \includegraphics[width=1\linewidth]{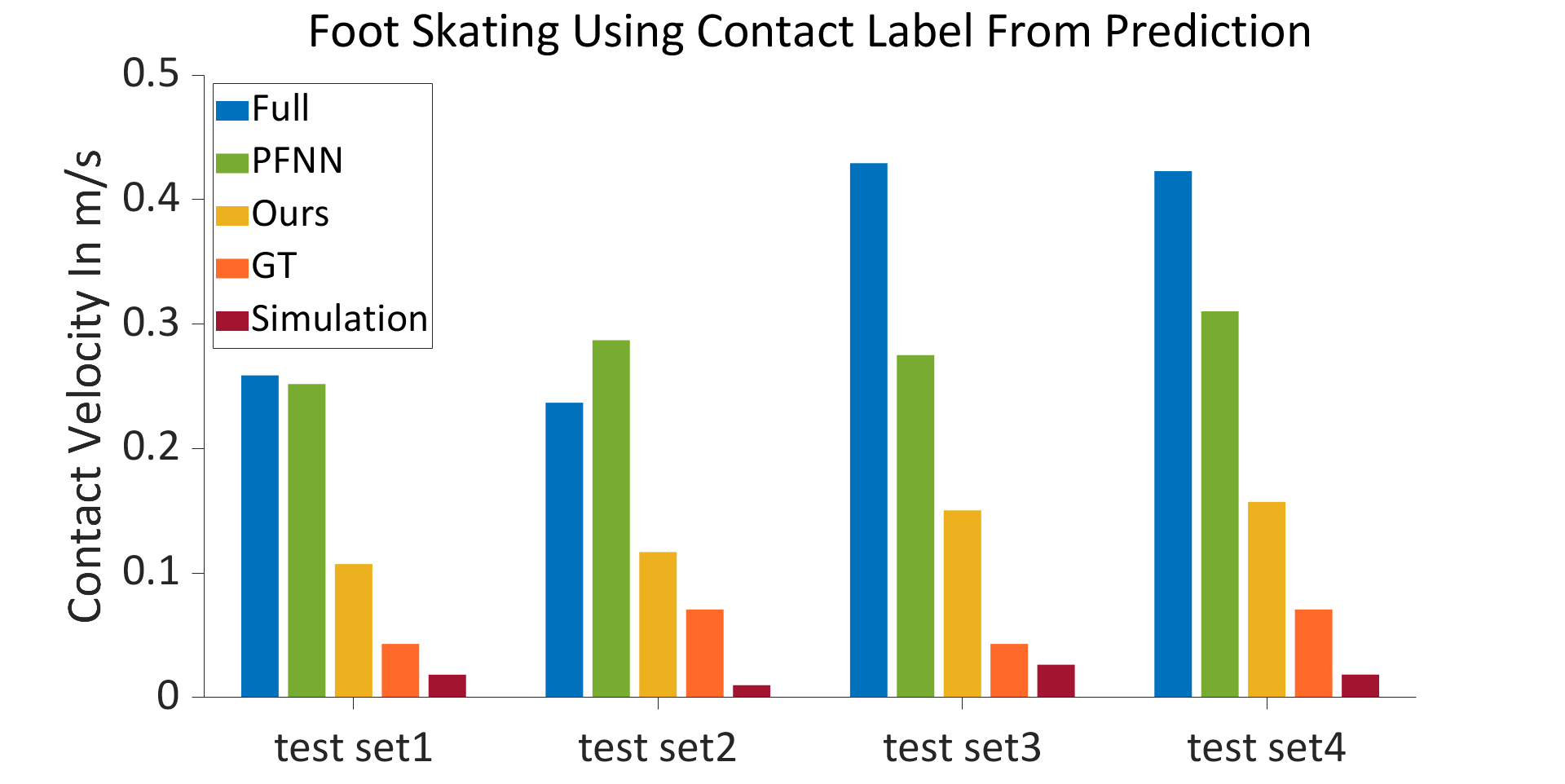}
    \caption{{Foot contact stability for different network architectures using contact labels predicted by the network.}}\label{fig:res:network_architectures}
\end{figure}

\begin{figure}[t]
    \centering
    \includegraphics[width=1\linewidth]{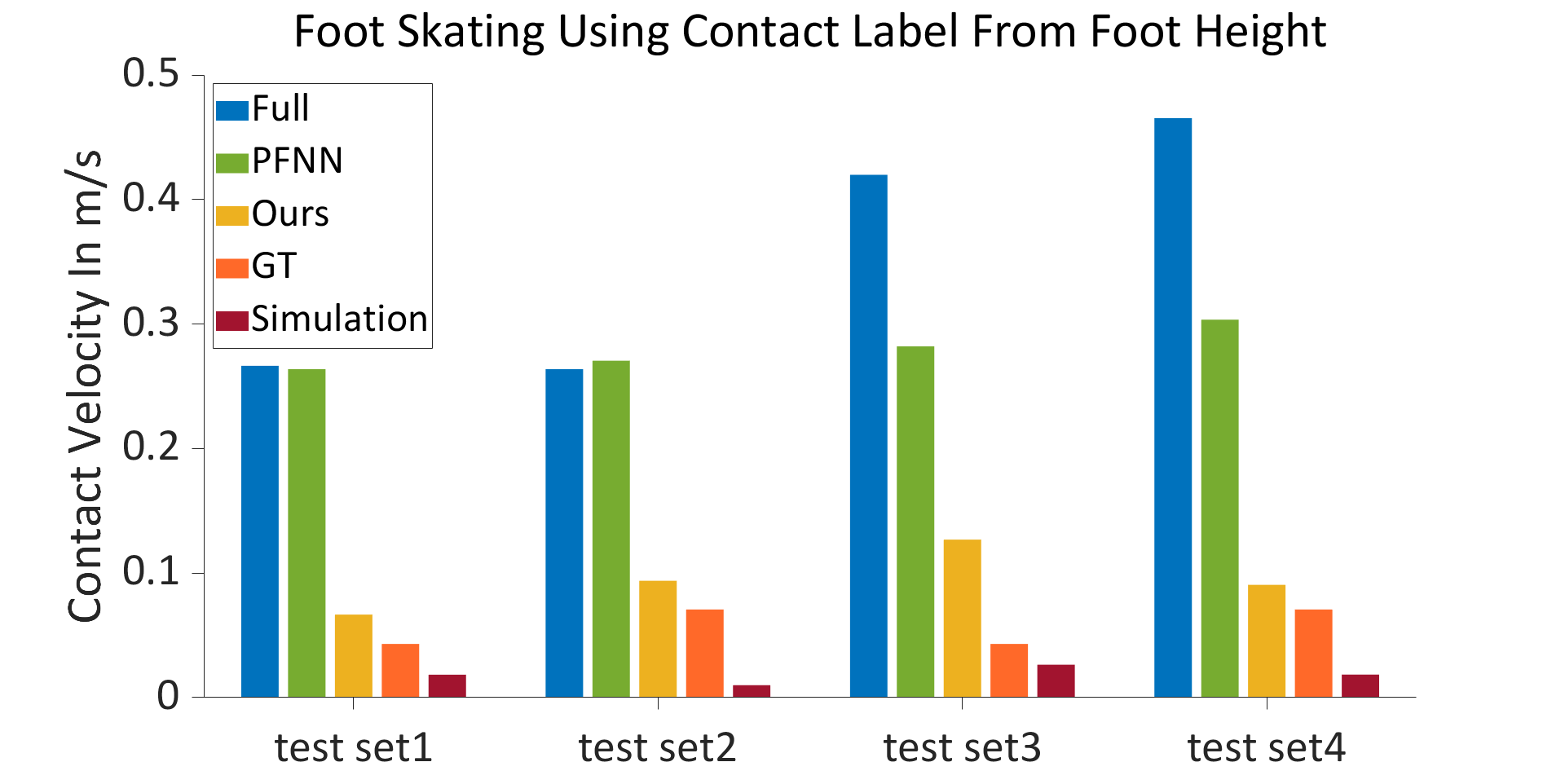}
    \caption{{Foot contact stability for different network architectures using contact labels calculated from foot height threshold.}}\label{fig:res:tracking accuracy}
\end{figure}

We train each network model on a small motion dataset of 36900 frames, 30fps. The assessment is performed on four test motion clips, labeled as test set 1, 2, 3 and 4, respectively. {Test set 1 is one of the training motions, where the character walks around randomly. Test set 2 is a similar motion to test set 1 but is not used in the training. Test set 3 is a synthesized motion, where the lower-body motion is taken from test set 1 but the upper-body motion are replaced by another motion clip with dramatic arm movement, which is not used in the training. Test set 4 is the test set 2 augmented in the same way as test set 3. All these test motion clips are clipped to 30 seconds long. The results are reported in Figure~\ref{fig:res:network_architectures} and \ref{fig:res:tracking accuracy}, which shows that our pose predictor generates motions with more stable foot contacts than the baselines, indicating that our decoupled network architecture is more robust to unseen upper-body input.}
As a reference, \fig\ref{fig:res:network_architectures} and \ref{fig:res:tracking accuracy} also show the performance of the normal mode of our system on the same test sets. It can be seen clearly that foot skating is effectively eliminated in the normal mode using the physics-based simulation. \lb{Note we also measure the stability of the foot contacts of the ground-truth motion in these figures, where the small amount of foot skating is due to the errors accumulated during motion capturing and retargeting.}
%

\subsection{Effectiveness of the Tracker Position Correction}
\label{ssec:Tracker_Position_Correction}
{We employ the Tracker Position Correction module in our system to improve the accuracy of tracking the positions of the hand-held trackers. To evaluate the effectiveness of this component, we employ a 30-second test sequence of the VR tracker signals recorded at 30\,fps, where the user walks around and waves {their} hand randomly. We then evaluate the tracking results with and without this component both visually and quantitatively.
While the generated lower-body movements are similar in both the settings, the generated hand positions match the corresponding input trackers better with this module on than turning it off. This can be seen quantitatively in Table \ref{tab:ablation}. Note that considering that our system bears a 0.1-second delay in the normal mode, we shift the generated motion forward by 0.1 seconds for a better evaluation of the accuracy. The results are shown under the \emph{Delay Removed} column of Table~\ref{tab:ablation}.
}

\begin{table}[t]
    \caption{Ablation study of the Tracker Position Correction component. 
    In the Delay Removed configuration, we shift the generated motion forward by 0.1 seconds to compensate the delay of our system, which provides a more accurate evaluation of the tracking results.
    }
    \centering
    \setlength{\tabcolsep}{1.5mm}{
    \begin{tabular}
    {c|c|r}
    \hline
    \hline
    {Tracker Position} &  Delay & \multicolumn{1}{c}{Positional}\\
        {Correction}  & Removed &  \multicolumn{1}{c}{Error (cm)} \\
    \hline
\ding{51} & \ding{55} & $9.95\pm 4.71$\\
\ding{55} & \ding{55} & $15.09\pm6.38$\\
\ding{51} & \ding{51} & $8.86\pm 3.56$\\
\ding{55} & \ding{51} & $12.21\pm5.84$\\
\hline
\hline

    \end{tabular}}
    \label{tab:ablation}
    \vspace{-10pt}
\end{table}

\subsection{Validation on Full-body Tracking Accuracy}
\lb{To further validate our system, we evaluate the accuracy of the  full-body tracking on a 30-second test sequence extracted from our mocap dataset. The test sequence, where the character walks around while waving hands randomly, is not used in the training process.
We use MPJPE~\cite{ionescu2013human3} as our evaluation metric, which is wildly used in the human pose estimation problem. We compute this metric based on the global positions of the joints in both the reconstructed motion and the reference.
Considering that the normal mode has a time delay of 0.1 seconds, we shift the generated motion forward by 0.1 seconds for better evaluation.   
The quantity results are shown in Table~\ref{tab:accuracy}.
Note that there are no explicit control of the character's lower body in the decoupled design of our pose predictor. The generated lower-body motion can be slightly different from the user's actual motion, causing a relatively large full-body tracking errors.}

\begin{table}[t]
    \caption{Tracking accuracy in different modes. The upper-body and lower-body accuracy are computed as the MPJPE of the joint groups $J_{\text{up}}$ and $J_{\text{lo}}$ discussed in the Section \ref{Sec:pos}, respectively. The generated motion is shifted forward by 0.1 seconds in the Delay Removed mode to allow a more accurate evaluation.
    }
    \centering
    \setlength{\tabcolsep}{1.5mm}{
    \begin{tabular}
    {c|c|c|r}
    \hline
    \hline
    {Tracking} & Mode &Delay & \multicolumn{1}{c}{MPJPE}\\
    {Accuracy} & Type & Removed &\multicolumn{1}{c}{(cm)} \\
    \hline
full-body & Normal & \ding{51}& $9.03\pm 4.10$\\
upper-body & Normal & \ding{51}&$7.66\pm4.09$\\
lower-body & Normal & \ding{51}&$12.52\pm 5.68$\\
full-body & Normal & \ding{55}& $11.81\pm 5.11$\\
upper-body & Normal & \ding{55}&$10.68\pm5.29$\\
lower-body & Normal & \ding{55}&$14.75\pm 6.38$\\
full-body & Direct &  - &$5.52\pm 4.73$\\
upper-body & Direct & - &$3.76\pm2.51$\\
lower-body & Direct & - &$7.88\pm 5.85$\\
\hline
\hline

    \end{tabular}}
    \label{tab:accuracy}
    \vspace{-10pt}
\end{table}

%% file: content/7_discussion.tex
\section{Discussion}
In this paper, we have presented a novel data-driven physics-based system for reconstructing full-body motions using a very sparse set of up to three VR trackers. Our prototype system can works with a typical VR system with its out-of-the-box functions to offer an enhanced immersive experience in virtual reality applications.
We have developed a full-body motion predictor module with decoupled upper-body and lower-body pose predictors to achieve a robust pose estimation, where the two components are combined via an aggregated representation of the state of the character. We find this network architecture outperforms the baseline methods that directly predict the full-body movement and is more robust with respect to unseen upper-body motions. 
We have trained a full-body control policy that controls a simulated character to mimic the user's action based on the prediction of the pose predictor module, which generates physically plausible motions with enriched details and allows the user's avatar to interact with the simulated environment and respond to perturbations.
To the best of our knowledge, we are the first deep-learning based three-point tracking system that achieves real-time tracking and simulation of full-body motions using such a small number of positional sensors.

Our method has several limitations. First, our system only predicts lower-body motions that are the most probable according to the transformations of the  upper-body VR trackers. Special lower-body motions, such as kicking and standing while swinging legs, are hard to predict using such limited information. It would be an interesting future work to include the information of the virtual/real environment as a part of the input to help determine the correct motion.

Second, while our decoupled pose predictor module is robust to unseen upper-body motions, the performance of the lower-body pose predictor is limited by the training data. {We find that our system can generate excessive foot skating in the direct mode when the user turns too fast, steps back quickly while turning, or acts with complex leg-crossing. The control policy and simulation in the normal mode of our system can remove the foot skating, but the character can perform unstably and may fall over,  
As a fallback, we can reset the simulated character to the state predicted by the pose predictor until the system resumes stable tracking.} Including additional data with dynamic motions would be helpful to mitigate this problem, while it remains a future work to develop a robust motion generator that can generalize to control signals with a different distribution from that of the training data.

\lb{Third, the normal mode of our system has a small amount of time delay, which may affect the user experience. While this time delay is intended in our system to allow additional information to be collected to track the user's motion accurately, reducing it to a more acceptable level will be a goal for future research.}

And lastly, our full-body pose-predictor and control policy are trained based on the same simulated character. While it is relatively easy to scale the input and output accordingly to support users of different heights, generalizing our system to support users with different body ratios and simulate non-human avatars remains a future work.

%% file: content/8_ack.tex
\section*{Acknowledgment}
This work is partially supported by the National Key R\&D 
Program of Science and Technology for Winter Olympics (No.2020YFF0304701) 
and the National Natural Science Foundation of China (No.61772499). 